\documentclass[journal]{IEEEtran}
\ifCLASSINFOpdf
\else
\fi

\usepackage{cite}
\usepackage{amsmath,amssymb,amsfonts}
\usepackage{algorithmic}
\usepackage{graphicx}
\usepackage{textcomp}
\usepackage{url}
\usepackage{url}
\usepackage{microtype}
\usepackage{hyperref}
\begin{document}
%
\title{A Literature Review on the Smart Wheelchair Systems}
%
%
%

\author{Yane Kim, Bharath Velamala, Youngseo Choi, Yujin Kim, Hyunkin Kim, Nishad Kulkarni, and Eung-Joo Lee}

\maketitle

\begin{abstract}
This study offers an in-depth analysis of smart wheelchair (SW) systems, charting their progression from early developments to future innovations. It delves into various Brain-Computer Interface (BCI) systems, including mu rhythm, event-related potential, and steady-state visual evoked potential. The paper addresses challenges in signal categorization, proposing the sparse Bayesian extreme learning machine as an innovative solution. Additionally, it explores the integration of emotional states in BCI systems, the application of alternative control methods such as EMG-based systems, and the deployment of intelligent adaptive interfaces utilizing recurrent quantum neural networks. A significant focus is on utilizing biosignals for motion detection, particularly in the development of head gesture-controlled wheelchairs, and applying deep learning techniques for signal analysis. The study also covers advancements in autonomous navigation, assistance, and mapping, emphasizing their importance in SW systems. The human aspect of SW interaction receives considerable attention, specifically in terms of privacy, physiological factors, and the refinement of control mechanisms. The paper acknowledges the commercial challenges faced, like the limitations of indoor usage and the necessity for user training. For future applications, the research explores the potential of autonomous systems adept at adapting to changing environments and user needs. This exploration includes reinforcement learning and various control methods, such as eye and voice control, to improve adaptability and interaction. The potential integration with smart home technologies, including advanced features such as robotic arms, is also considered, aiming to further enhance user accessibility and independence. Ultimately, this study seeks to provide a thorough overview of SW systems, presenting extensive research to detail their historical evolution, current state, and future prospects.
\end{abstract}


%
\IEEEpeerreviewmaketitle

\section{Introduction}
\label{sec:introduction}
Smart wheelchairs (SW) stand as a notable technical innovation, significantly enhancing mobility and autonomy for individuals with various physical limitations. These devices integrate advanced software and hardware, such as sensors, artificial intelligence (AI), and user-friendly interfaces, to safely and efficiently navigate complex environments. This technological convergence not only enhances user capabilities in everyday tasks but also promotes greater accessibility and independence.

This review paper investigates the extensive research on SW systems, providing a detailed overview of their historical evolution, current developments, and potential future directions. A primary focus is the advancements in Brain-Computer Interface (BCI) technology, particularly its role in improving SW control to augment independence for users with severe motor disabilities. The paper examines diverse BCI systems, including mu rhythm, event-related potential, and steady-state visual evoked potential, which enable the conversion of cognitive processes into actionable commands. Despite progress, challenges such as EEG data fluctuations and noise persist. Current research efforts include the exploration of methods like sparse Bayesian extreme learning machines and particle swarm optimization-based rough set feature selection to enhance accuracy.

The paper also highlights the ongoing integration of emotional state evaluation in BCI, aiming to boost user engagement. There is an active development of alternative wireless control methods, such as steady-state visual-evoked potential and EMG-based control, to improve both efficacy and convenience of use. Yet, executing multiple commands through basic BCI devices remains a challenge. Innovations like intelligent, adaptive user interfaces utilizing recurrent quantum neural networks are emerging as potential solutions.

Further, the review explores the use of biosignals for motion recognition, with a special focus on head gesture-based control systems for wheelchairs. It discusses strategies aimed at minimizing continuous user input by adopting learned operating modes. Recent studies have been employing advanced deep learning techniques, like Recurrent Neural Networks (RNNs) and Long Short-Term Memory (LSTM) networks, to more accurately categorize signals and enhance SW performance. The paper also sheds light on advancements in autonomous tracking, navigational support, localization, and mapping, showcasing various sensor and computational architectures.

Moreover, the human aspect is critical in SW systems, necessitating fluid interaction between the user and the wheelchair. This includes considerations of privacy, physiological factors, and the refinement of input controls. The complexity and context-dependent nature of control inputs pose significant challenges. The review acknowledges the commercialization efforts by entities such as Smile Rehab, while pointing out the constraints associated with indoor use and the need for training.

Looking to the future, the paper anticipates advancements in autonomous navigation and interaction models, adapting to dynamic environmental conditions and user preferences. Research is ongoing in areas, such as reinforcement learning and varied control modes, including eye and voice control, to further refine the interaction between users and their wheelchairs. In addition, the burgeoning field of smart homes, integrating features such as robotic arms and gaze-driven interfaces, is discussed, highlighting its potential to bolster accessibility and foster greater independence for users.

In this paper, the sections are summarized as follows: Section 3 covers the historical development of powered wheelchairs, including their key components and utility. In Section 4, we then summarize the current state of wheelchair development, focusing on input modes, operating mechanisms, and human factors. Section 5 provides a summary of the future directions for smart wheelchairs, discussing aspects such as autonomous navigation, Human-Smart Wheelchair Interaction models, and smart home applications, along with other potential practical scenarios.

\section{Method}
\subsection{Search Strategy and Criteria}
In conducting this survey report, the search engines Google and Google Scholar were utilized. For this investigation, papers not accessible via Google Scholar were excluded. Furthermore, our primary emphasis is on the past decade, during which we have analyzed advancements in both hardware and software aspects.



\section{Past: Powered wheelchair}
\subsection{History}
Experiments with powered wheelchairs have been conducted for a long time to improve the lives of individuals with mobility impairments. The first fully functional motorized wheelchair emerged in London in 1916. Subsequent modifications to this wheelchair model by various researchers included the addition of movable arms and a footrest. Later, the introduction of a foldable design addressed the previous models' issues of being heavy and cumbersome\cite{past_PW1}. However, the modern powered wheelchair, as we know it today, was developed in 1953, largely due to the efforts of George Klein, an engineer at the National Research Council of Canada. Inspired by the need to assist World War II veterans who were quadriplegic, Klein designed a lightweight, portable, and easy-to-control wheelchair\cite{past_PW2}. Powered by two batteries, it had a range of approximately 20 miles and a top speed of 2.5 mph, features Klein deemed optimal for indoor use. Klein's invention significantly enhanced mobility for individuals with disabilities, providing them, for the first time, a means to achieve independent mobility and fully engage in society\cite{past_PW3}.

The powered wheelchair quickly gained popularity and became an essential tool for millions of people worldwide. Recognizing its potential for a wide range of disabilities beyond immobility, mass production began in 1956, led by companies such as Everest \& Jennings and the American Wheelchair Company\cite{past_PW4}.

\subsection{Key Components}
Initially, powered wheelchairs were developed with a focus on heavy hardware. However, over the following decades, technological advancements led to the integration of sensors, smart controllers, software, and machine learning. The basic components of powered wheelchairs include:

\begin{itemize}
  \item Seating: A key goal for researchers was to develop seating systems in power wheelchairs (PW) that provided optimal comfort. This led to the use of materials like foam and gel in cushions, enhancing comfort and preventing pressure sores — a vital consideration given the long periods users spend in their wheelchairs. Additional features included ergonomically designed backrests, footrests for proper foot placement, and armrests for hand support, collectively forming a recliner system that improves overall comfort and addresses specific ergonomic needs\cite{past_PW5}.
  \item Drive system: Power wheelchairs (PW) are equipped with a power base that includes drive wheels. These wheels can be front-wheel drive, center-wheel drive, or rear-wheel drive, depending on the model. Front-wheel drive is often preferred in many designs due to its superior maneuverability\cite{past_PW2}.
  \item Batteries: Batteries are essential for powering the wheelchair and sustaining its functionality. Given their finite capacity, users often need to carry an extra battery for longer trips. Air travel presents additional challenges, as batteries must be removed during flights, leading to the adoption of compatible dry cell batteries.
  \item Controller: The motorized controller acts as the interface between the wheelchair and the user. It usually features various types of joysticks, such as hand, chin, and head joysticks, enabling users to navigate with greater independence and less physical effort\cite{past_PW6}.
  \item Chassis: Different chassis designs are used to meet various needs, including foldability and all-terrain navigation. These designs aim to enhance user flexibility, allowing comfortable access to diverse locations due to the foldable nature of the chassis\cite{past_PW7}.
  \item Smart Controller: The integration of Machine Learning technology, cameras, and sensors, as well as research into gesture-controlled and brain-controlled mechanisms, are advancements in the field. These features aim to enhance the wheelchair's functionality and user experience.
\end{itemize}

Several other components are crucial for the optimal functioning of powered wheelchairs. These include materials that withstand diverse terrains, a precise braking system for user and bystander safety, the integration of communication devices, and the use of various sensors.

\subsection{Different Application for Smart Wheelchair}
Individuals with diverse disabilities are inclined to go for different control methods, posture settings, and destinations based on their unique needs and personal activities. Users with significant physical impairments necessitated comprehensive postural support to ensure proper alignment of the pelvis and spine, prevent deformities, and maintain a stable sitting balance. In the development and commercialization of previous power wheelchairs, challenges arose in selecting dimensions suitable for indoor use.

Historically, previous research has effectively accomplished initial technological enhancements that assist wheelchair users in their daily routines. However, there has been limited inclusion of assistive technology within the wheelchair with smart or autonomous capabilities, aligning with the broad aim of research focused on integrating smart technology into power wheelchairs. Many prototypes of Smart Wheelchairs have been developed and brought into market since the 1990s. Challenges persists including the balancing of autonomy between user control and system control. Smart Wheelchairs are a promising research area to help the mobility-impaired with many advancements in the past\cite{past_PW8}.

\section{Present}
\subsection{Input Mode}
The Brain-Computer Interface (BCI) is a promising technique that utilizes brainwave signals instead of motor control to steer SW. This innovative approach provides the means to interact with the physical world mechanically for people with severe motor impairment, e.g. due to neurodegenerative disease or impaired peripheral nervous systems (Holz et al., 2012; Kaufmann et al., 2013). By enabling users with any degree of disability to operate the SW, the BCI technique fosters autonomy, enriching their engagement with the surrounding environment (Kannan et al., 2013).

\subsubsection{Neural Responses Used in BCI Systems}
The BCI system typically utilizes three types of neural responses: mu rhythm, event-related potential, and steady-state visual evoked potential. Mu rhythm, spontaneous neural response, occurs during the motor imagery (MI) task (Jin et al., 2020). MI denotes the cognitive preparation of a motor activity devoid of physical muscle engagement. A BCI wheelchair based on MI classifies electroencephalogram (EEG) rhythms made by a patient's mental processes into operational commands. One of the primary challenges in BCI application pertains to accurately categorizing MI due to the variability of EEG signals among users (Lotte et al., 2007) and background noises possibly caused by EOG and EMG (Park et al., 2013). Moreover, the extracted EEG MI features exhibit high dimensionality, which reduces the accuracy of classification. To increase the accuracy of classification for high performance, researchers have been introducing new methodologies and algorithms, such as sparse Bayesian extreme learning machines (Jin et al., 2020). In a notable study, Kumar and Inbarani combined particle swarm optimization (PSO)-based rough set feature selection techniques with a novel neighborhood rough set classifier (NRSC). Their findings demonstrated the superiority of these approaches over the BCI Competition IV Dataset IIa (2017). Carlson and Millan (2013) employed an asynchronous MI approach within a shared control architecture. The shared controller integrated both the user's input and intelligence, which aligns with the two-layer collaborative approach proposed by Lopes et al. (2013). The collaborative approach involves two distinct layers. The first layer determines whether to enable a command. The second layer makes the final decision while taking into account the user's competence, which was evaluated by comparing the user’s judgments and the expected steering command. This design provides a higher degree of flexibility and authority to the users over the actual trajectories of the wheelchair.

\subsubsection{Collaborative and Emotional BCI} 
Fattouh et al. (2013)  introduced a system that assesses users' emotional state while executing commands, determining whether to halt the command or proceed. The authors propose that this emotional BCI control system addresses a limitation of traditional BCIs, as it doesn't demand a high level of focus on external stimuli.
Using steady state visual-evoked potential (SSVEP) or P300, a resonance in the visual cortex during focused attention on a light source flickering above 4Hz, based BCI, users with not much training can utilize the SW more easily compared to non-visual evoked based BCI systems (Regan, 1989). Diez et al. (2013) conducted an evaluation suggesting the utilization of high-frequency stimuli for SSVEP-based BCI due to its benefits, including lower error rates and reduced visual fatigue due to flickering displays. Limitations of using SSVEPs or visually elicited event-related potentials (ERPs) as input signals include the need for a display in the user's visual field, hindering observation for those with severe impairment, inability to observe surroundings during target selection due to visual focus, and potential efficacy reduction under changing light settings (Kaufmann et al., 2014). Considering these factors, Kaufmann and researchers demonstrated the feasibility of tactually evoked ERPs by using tactile vibration units called tactors.
EMG-based control systems are gaining popularity for hands-free operation. However, wired connections for signal acquisition pose a drawback. To overcome this, researchers proposed wireless EMG control. Champaty et al. (2014) introduced wearable signal acquisition, transmitting control signals wirelessly to the wheelchair's servo-motors for improved convenience and mobility. 
Despite the advantages of BCI for enabling communication with SW for users with impaired movement, a simple 2-class BCI system's output signal has limitations, making it challenging to execute multiple motion commands. Although using a multiple-class BCI can address this, it reduces classification accuracy and necessitates a complex protocol (Obermaier et al., 2001; Kronegg et al., 2007). To tackle these challenges, Gandhi et al. (2014) proposed an intelligent adaptive user interface (iAUI) within the adaptive shared control BCI system. They employed the synchronous mode of BCI operation and utilized a recurrent quantum neural network (RQNN) method to filter EEG signals. The implementation of iAUI resulted in an improved information transfer rate, as it offers an updated prioritized list of selection options (Gandhi et al., 2014).

\subsubsection{Emerging Trends in BCI: Visual, Tactile, and Wireless Technologies} 
Biosignals, including Electroencephalogram (EEG), Electrooculogram (EOG), and Electromyogram (EMG), offer notable advantages as they remain viable even in cases of complete paralysis in patients (Adebayo \& Ajayi, 2021). Motion recognition extends beyond the realm of limb-based movements to encompass even head gestures, which offers substantial benefits for quadriplegic patients who lack control over their extremities. Pajkanović and Dokić (2013) introduced a microcontroller system with an accelerometer that captures force-related information emanating from head movements, enabling the microcontroller to compute data. This data ultimately steers a mechanical actuator, thereby controlling the joystick of a wheelchair and directing it in accordance with the user's head movements.for head gesture data collection. Machangpa and Chingtham (2018) enhanced this technology with gyroscope data and advanced algorithms for obstacle avoidance. This innovation enhances mobility and exemplifies the field's progress.

According to a review conducted by Leaman and La (2017), it is recommended to prioritize the reduction of users' reliance on continuous input commands by implementing operating modes that can be educated based on users' everyday activities.

\subsection{Operating Mode}

\subsubsection{Machine Learning}
In the current research, deep learning techniques, such as Recurrent Neural Networks (RNNs), are being widely employed to handle time series data while preserving interdependencies among data inputs. A recent study by Adebayo and Ajayi (2021) introduces a hand gesture recognition-based control system for smart wheelchairs. The study utilizes an RNN architecture known as Long Short-Term Memory (LSTM) to classify electromyography (EMG) effectively into specific hand gesture categories. By combining biosignal control systems with Artificial Neural Networks (ANNs), this approach demonstrates exceptional performance and represents the current state-of-the-art for real-time applications, including computer vision and natural language processing tasks (Adebayo \& Ajayi, 2021).

\subsubsection{Following}
To alleviate the burden on caregivers and enhance communication between wheelchair users and their companions, there is a notable demand for the development of SW systems capable of autonomously following the companion. These "following" wheelchairs employ Adaptive Cruise Control (ACC) technology to facilitate platoon driving. These platoon driving systems have been devised not only for navigating straight paths but also for maneuvering through narrow corridors. The platoon control for narrow space is achieved through the utilization of inter-vehicle distance control based on ACC, coupled with lateral control following the wheel track of the preceding wheelchair, as previously discussed by Sugano et al. in 2014.

In the pursuit of effective companion tracking, researchers have employed Kobayashi's tracking algorithm from 2012, which harnesses Laser Range Sensors to precisely monitor the position of the companion. Furthermore, Murakami et al. in 2014 introduced a model that ensures the wheelchair maintains a side-by-side orientation with the companion, even when the user's destination is uncertain, as in the case of leisurely activities like window-shopping. Using linear extrapolation of velocity, they came up with a system that lets the wheelchair move along with the predicted path of the companion.

In addition to these developments, Suzuki and colleagues in 2014 proposed a multi-wheelchair robot system that achieves synchronization of movements among several wheelchairs and their respective companions. Building upon Kobayashi's algorithm from 2012, their system incorporates sensor poles, resulting in less occluded observations when tracking multiple companions. Notably, their system stands out for its capacity to construct an environmental map, allowing the wheelchair to estimate its own position and thereby adapt its formation for improved communication.

\subsubsection{Navigational Assistance}
Numerous different versions of smart wheelchairs incorporate different sensors and computing structures on a power wheelchair or a seat on a robot for semi-autonomous or fully autonomous navigation. Hartman et al 2019 [] incorporated fundamental functions for smart wheelchairs, including path finding, motion control, localization, perception and computer cluster network to process real-time data. 
\begin{itemize}
  \item Perception: LRF and RGB camera data are processed and transferred to the computers designated for the perception task to present the terrain obstacle data. 
  \item Path finding: perception data are transferred from perception-designated computers to the cognition computer over the Ethernet LAN
  \item Computer cluster network: it consists of four small form factor (SFF) desktop computers that are interconnected over a Local Area Network (LAN). The cluster is specifically configured as an Internet Protocol version 4 (IPv4) class C private network. Special data processing hardware allows for real-time processing of information from the sensors for the most efficient autonomous navigation. A computer cluster network allows for cloud computing, facilitating the parallel processing of a larger amount of data, which is crucial for evolving sensors and more complicated smart wheelchair systems. 

Simpler smart wheelchair designs incorporate a navigation subsystem, location monitoring subsystem, voice guidance subsystem, and obstacle detection subsystem (Rathore et al 2014). 
  \item Navigation subsystem: composed of an accelerometer and magnetometer modules, allows for movement towards any desired direction. The navigation pad can be tilted in any direction along the ~\textit{x} and \textit{y} axes, allowing for precise movement. The magnetometer allows for directional indications as the vibration pad attached to the chair with four vibration motors gets activated based on the direction the wheelchair is moving.
  \item Location monitoring subsystem: RFID tags are placed in the building of the wheelchair with RFID receiver attached to the wheelchair to receive signals from the RFID tags and indicate the location of the wheelchair.
  \item  Voice guidance subsystem: A MP3 player is attached to the microcontroller, so when an RFID tag is encountered by the wheelchair, the unique ID number is received by the microcontroller. Then, the microcontroller sends signals to the MP3 player to play the pre-recorded audio clip to the driver.
  \item Obstacle detection subsystem: ultrasonic sensors emit sound waves in inaudible frequencies for the human ears that get reflected back after hitting an object. The reflected waves are received by the sensors to calculate the total return time and the distance from the object.
\end{itemize}

\subsubsection{Localization and mapping}

With the development of technology, the basic definition of what makes a smart wheelchair ‘smart’ has also evolved. Older papers in their definition use technologies such as ultrasonic sensors coupled with a control system to include obstacle avoidance in their designs. Modern smart wheelchairs can operate at a higher level of autonomy. In a known environment, it would be easy for a wheelchair to navigate from point A to point B, such that a global plan would be calculated\cite{apriori} with only an additional need to perform obstacle avoidance along the way for dynamic obstacles\cite{comfortable}. This means that we have our autonomous agent knows it's location in the known map. For autonomous movement and navigation through an unknown environment, a wheelchair must first tackle the problem of mapping its environment and then also simultaneously localizing itself within that environment. The procedure to do this is called SLAM, or Simultaneous Localization and Mapping.

The most common method of doing SLAM in real-time is to use LIDARs, i.e., Light Detection and Ranging. The advantage of using LIDARs is that they can provide highly accurate real time depth maps of a plane of observation. This is achieved using infrared radiation from special emitters and receivers on the LIDAR. However, the information is only available for a single 2D plane.

Alternatively, RGB cameras can be used to perform Visual SLAM. These methods are a subset of the Structure from Motion methods used to reconstruct an environment. The greatest advantage of using RGB camera is that of the ability to deploy additional algorithms like Object detection using YOLOv5 \cite{YOLOv5} which can be used for additional obstacle avoidance\cite{Kinect Wheelchair} thus assisting navigation. Visual SLAM can be further sub-divided into monocular, stereo, and RGB-D SLAM. Monocular SLAM extracts depth by tracking changes between subsequent frames. Stereo vision uses multiple views of the same scene. Finally, RGB-D cameras use RGB cameras to extract 2D views and fuse them with Depth information extracted using infrared sensors, creating a fully 3D view in a single frame. An extension of Visual SLAM is visual Inertial Odometry which extends SLAM to also find the path taken by the camera. 

One of the challenges of SLAM is loop detection and loop closure, this can be done by various methods like Monte Carlo\cite{ros_lidar}, \cite{{monte_carlo}},occupancy grid maps\cite{grid map}, EKF SLAM \cite{urban}, UKF SLAM\cite{ukf}, etc. Another way to differentiate the versions of SLAM is on the basis of hardware. Some use LIDARs\cite{ros_lidar}, while others may use stereoscopic cameras, RGB cameras, or RGB-D cameras\cite{astar_ros}. Juneja et. al. conducted a study to compare various methods of SLAM for autonomous wheelchairs\cite{slam_review}.

In recent years, radiance field methods have shown great promise in real time structural reconstruction. These methods have been extended to SLAM and have even found commercial use. Since these methods are highly accurate in reconstructing the environment, they may be used in autonomous wheelchairs in the future.

A combination of deep-learning algorithms and measurement sensors were used with a specific dataset for object detection, localization, and tracking from a study by Lecronsnier at el 2021. They used a convolutional neural network (CNN) that utilized deep-learning algorithms to process images acquired on the D435 camera, a RGB depth image camera. The objects (doors and wheelchairs) were located by a Vicon system, and the distance between the wheelchair and the objects was calculated. Then the object was tracked by adding the 3D object to the semantic map with the odometer data from the T265 camera.

\subsection{Human Factor}
The smart wheelchair represents an assistive technology that combines a power wheelchair, which utilizes motors for movement, with an integrated computer system and sensors. The inclusion of advanced machine learning technology facilitates interaction between the wheelchair system and the user. Sensors collect data from the surroundings, which is then processed to enable the wheelchair to move based on decisive steps. The smart wheelchair can be controlled manually by the user, autonomously by the system, or through a shared control mechanism between the user and the system.

Despite advancements in machine learning technology, achieving a functional smart wheelchair requires addressing several challenges related to human factors. These include establishing dependable and seamless interaction between the smart wheelchair and its user, addressing privacy concerns, understanding the physiological aspects of the user, refining input controls, and exploring the interaction between the wheelchair and the external environment. Ongoing research on various types of smart wheelchairs continues to prioritize human factors\cite{human_factors1}.

With the introduction of numerous input controls in smart wheelchairs, choosing the best one, considering both the ideal outcome and user preferences, has become complicated. Controls such as language detection, facial detection, hand gestures, voice recognition, touch screens, and eye tracking could be used for operating the smart wheelchair. Research on shared-control smart wheelchairs, where users and the wheelchair system interact to achieve a user-decided outcome based on information collected by sensors, is ongoing\cite{human_factors1}. However, some controls may be inaccessible to users with certain disabilities affecting hand movement or voice recognition. Additionally, some tools may only be functional indoors, as bright daylight or darkness can impair certain functionalities. An example of overcoming these limitations is the iChair, which uses standard power wheelchairs equipped with LEDs for lighting, a 3D scanner, an HD camera, a head mouse, a reflective dot, a head tracking mouse for input controls, and a laptop to manage the software. The development of an EEG machine to translate brain waves into inputs is underway but has not yet reached feasibility\cite{human_factors2}.

Prioritizing human factors is essential in smart wheelchair research and development. The following subsections will focus on crucial components that need to be addressed in human factors as part of smart wheelchair research.

\subsubsection{User and Robot Interaction}
The human-robot interaction involves numerous interactions between the user and the smart wheelchair. Many different inputs could be utilized and specified for different users who have different needs. Users who have disabilities moving their hands and such may require a head tracking harness or eye tracking software, whereas those who don't have such disabilities may just need a way to get inputs and a joystick for maneuvering the smart wheel chair. Users do have preferences owing to their disability and their personalities regarding the type of inputs they want and the manner in which they prefer to receive them from the smart wheelchair system\cite{human_factors4}.

There have been multiple studies going on on different ways of delivering the inputs or feedback generated by the Smart Wheelchair system. Some of these ways include providing inputs through visible feedback on a screen, haptic feedback, preferably wearable vibrotactile feedback, a simulation of spring effect, a haptic feedback through the use of joystick, and auditory feedback to the users. Research is going on using a combination of the aforementioned ways of providing feedback or input. One thing that needs to be highlighted is that sticking to one type of interaction cannot be scaled to most of the users; as long as the  Smart Wheelchair is customizable, the users can choose their preferred mode of interaction.\cite{human_factors5}

Other concerns may also arise from the human-robot interaction, such as privacy, ethical, and moral concerns that may arise due to cameras being used for tracking where consent may not be given to surrounding individuals for tracking or to the user for intimate moments\cite{human_factors6}. The manufacturers or researchers should be able to draw a line when to collect the data by vigorous testing of the surroundings, either through location-based knowledge or visual-based knowledge. And the users should also be informed about the SW system; they should be able to turn off or turn on the data tracking part of the system, either the cameras or other sensors\cite{human_factors7}.
\subsubsection{Crowd and Robot Interaction}
Crowd and Robot Interaction seem to be receiving comparatively less attention in the ongoing research of SW. It is important to note that SW users frequently experience a sense of obstruction in regular, crowded situations. It would be beneficial to inform the crowd about the direction of the wheelchair's movement and its destination. The smart wheel chair signals the external crowd through some medium about the movement of the smart wheelchair, ensuring smoother navigation without any collisions or obstructions. Various methods can be employed to signal the external crowd, including visual and auditory means. Visual means include displaying information of the direction or route on a screen that can be seen by the crowd, or directional arrows displayed on the road by LEDs can be a good and interpretable way for the crowd\cite{human_factors5}.

Though auditory means make sense to use in these situations, the users of Smart Wheelchairs do not want to be announced whenever they are coming or going through a crowd. But when there is a busy crowd or loud surroundings, an auditory signal is useful. It is recommended to have whatever means possible integrated into the system so that the user can choose whatever suits the situation.
\subsubsection{Adaptability for the users}
With the growing integration of technology into Smart Wheelchairs, the convenience of understanding how to use them by common people is more burdensome for those who are already disabled. Users need to learn how Smart Wheelchairs work and what each control means, allowing them to maneuver seamlessly\cite{human_factors9}. This learning process can be used as a feedback session for manufacturers and researchers so that Smart Wheelchair can be developed based on the preferences set by the user, which can be customized. Since the user needs to spend most of their time on Smart Wheelchair itself, they must be comfortable sitting on it. It is applicable when the terrain is not smooth, since the users feel the impact of the irregular terrain. Users with disabilities shouldn't receive any physiological impact\cite{human_factors10}.

There are also safety concerns that come with human-robot interaction that may arise if the inputs do not match the outputs that the user and the developer of the smart wheel chair desire. Glitches and damage to the system may leave the user stranded if they do not have help nearby. Testing of smart wheelchairs involves human subjects that have high standards for smart wheel chairs to be able to pass. With such strict guidelines, there is currently only one commercially available smart wheel chair company known as Smile Rehab\cite{human_factors8}. Licensing users by the government to use Smart Wheelchairs making sure that the 
\subsubsection{Social Awareness}
Users of the smart wheelchair tend to feel a deep connection with the system, considering it an integral part of themselves. They desire that the incorporation of the smart wheelchair into their daily lives does not impact their interactions with others. It was known through a survey in a research paper that Smart Wheelchair users usually think through how they are perceived among the crowd. Smart Wheelchair users don't usually want them to shine a spotlight in a crowded scenario while they are passing by. Any kind of auditory interaction with the user will definitely put them in the spotlight, which defeats the purpose unless it is really necessary owing to the disability of the user\cite{human_factors5}.

There are situations where users could not properly engage in a conversation with others due to the obstruction of a wheelchair body, like a laptop or a screen. This limitation can hamper their ability to fully engage in social interactions and prompt a need for solutions that accommodate seamless communication. This presents us with limited opportunities to choose how the system can interact with the user. Providing flexibility to the Smart Wheelchair such that users can customize as they intend based on their needs and activities they would like to perform
\subsubsection{Commercialization}
Smile Rehab is currently the only commercially available smart wheel chair that can only be used indoors. They require the user to be trained for the smart wheel chair with progression levels that first require the use of their training device called the Drivedeck. The users are trained using this device for the usage of their Smile Smart System Powerchairs, which utilize software that comes with anti-collision sensors, voice confirmations, and pre-determined tracking with speed and motion controls. They also utilize tapes that are set up in certain locations for the detection software on the smart wheel chairs\cite{smile_rehab}.

For commercialization purposes and for the continuous well-being of the users, vitals monitoring information can also be added to the user's screen as a dashboard. This allows users to monitor themselves rather than depend on others\cite{human_factors11}.

\section{Future}
The future of SWs prioritizes giving users full autonomy, which further increases their confidence in the SW and new features. A study conducted by Worcester Polytechnic Institute (WPI) shows that the majority of SW users are open to a robot that assists them with performing tasks in an environment with other people (Padir et al, 2014). Named "Co-robots", these robots provide the user with independent mobility. Alongside autonomy, SW research must prioritize the safety and reliability of their wheelchairs. In this section, we will explore the factors that must be placed at the forefront of SW research and development.

\subsection{Autonomous Navigation}
An autonomous driving system is expected to be established in the system of SWs as one of the future bio-engineering techniques for the SW. For this, it is critical to analyze graphs describing tracking errors, system architecture, planning paths, planned steering angles, and controller performance in order to apply the self-driving system to the SW \cite{b1}.

Future SW users will be able to steer their wheelchair along pre-programmed pathways, avoid obstacles in their navigation paths, and easily pass through doors. Additionally, the SW will be established as a “customized SW”, which allows the user to build an individualized profile. This profile contains the user's preferred input method, rate of wheeling and turning, amount of verbal feedback, frequently used pathways, etc.\cite{b13}. Furthermore, the most essential component of the self-driving SW is its ability to modify its navigation according to the changes detected within the user's environment. One study evaluates the efficiency of semi-autonomous control to allow the reasoning of navigation. It adopts an extension of a Vector Field Histogram (VFH+), using the 2D Cartesian histogram. This histogram is built around the wheelchair in a square shape, and each square grid is transformed into the Polar Obstacle Density (POD), a 1D histogram representation. As a result, the wheelchair can avoid collision in semi-autonomous modes as such to reduce the user’s continuous monitoring of surroundings while maneuvering \cite{b2}.

However, because the environment continuously changes within one's surroundings, it is evaluated as useless to utilize auto-navigation permanently to SW. In contrary to deliberate planners who perform expensive and limited path planning for robotic automation, local planners provide cheap and broad planning pathways based on local surroundings \cite{b4}. Online calibration and sensor sources can also help to resolve this limitation. A variety of sensors can detect the movement and appropriate steering angle of a 
SW \cite{b3}. Notably, a newly developed sensor can detect the flatness of the ground and the presence of obstacles on the ground, which can be applied to future SW models \cite{b3}. Even voice recognition is expected to be invented to drive the wheelchair and provide seat adjustment through voice commands \cite{b7}. Additionally, an omnidirectional-wheeled-robot, which uses a decentralized algorithm to control its movement, has been successfully developed to overcome the complexity and difficulty of the conventional methods of controlling SW movement. For example, each algorithm analyzes the potential direction of each pathway in which the wheelchair can move \cite{b5}.   

Since there exists a precedence for autonomous science technologies such as autonomous robots and cars, the autonomous SW is a feasible invention as long as it has a navigation algorithm that allows it to adapt to the surrounding environment \cite{b6}.

\subsection{Human-Smart Wheelchair Interaction Model}
Establishing the human and SW interaction model is essential while we attempt to create a new generation of SWs. To develop the model, sensor feedback, as well as a reinforcement learning technique, can be used to build the algorithm \cite{b12}. Reinforcement learning is based on the idea that the agent interacts with the environment at discrete time steps. So, the agent uses a scalar value reward from the environment to the agent. Especially, reinforcement learning (RL) is considered a long-term optimization method for a shared control system in which the user and SW collaborate together. The shared control system focuses on the extent of physical function of users and their driving motivation. By investigating the operating features and lifestyle of target users, the shared control system can be trained to consider the operational ability of users \cite{b8}. Currently, an interface that is controlled by eye and voice recognition has been designed to assist SW users without inconvenience. The interface, which contains an eye control mode and a webcam that captures the real-time eye movement of the user to generate the direction signal, is able to operate the wheelchair in a specific direction. 

Additionally, a multi-rate AMR voice app is used for voice control mode by commanding four kinds of signals; ON (forward), left, right, and stop. Arduino UNO is the main controller, and other controllers such as the microphone, HC-05 Bluetooth module, and AMR voice recognition, are used for sub-controller.

Furthermore, a finger-gesture-controlled wheelchair has been invented for SW users. The control system for finger gesture control undergoes an algorithm of hand detection, hand tracking, gesture recognition, and operation of the wheelchair. This algorithm has already passed an accuracy test, yielding over 97 percent accuracy in factors including obstacle detection, fall detection, and an emergency messaging system through use of an SMS server. Additionally, this system has the potential to enhance the quality of gesture control \cite{b9}.

\subsection{Smart Wheelchair with Smart Home}
The creation of a smart wheelchair with a robotic arm to assist an individual living within a smart home is gaining attention among SW designers. The smart home is a new trend of housing that supports automation. Wi-Fi is the main remote monitoring device, and this device is an essential component of IoT, which are monitored and controlled remotely over the internet. Smart homes using smart technology can be used to supplement independent living by increasing accessibility for the daily activities of wheelchair users within their residential environment. Therefore, the smart home is a living space that supports independent living and provides various technologies to maintain health (Myungjun et al, 2016). Since there are many approaches to combining the smart wheelchair into a smart home, a robotic arm and retractable roof are being invented so that they can be embedded with the SW. SWs with a retractable roof and movable arms can guide users to perform daily activities within the home. They can also provide extra security while users are outside driving SWs by communicating remotely with the smart home system. Wästlund and other smart engineering researchers have developed the EyeGo System, which can be fitted to a SW to create a gaze-driven power wheelchair. With the EyeGo System, individuals with disabilities can steer their electric wheelchair indoors using their eye movements. The system is composed of a gaze-driven user interface, an intelligent junction box, and a navigation support system \cite{b11}. When the navigation support system is successfully compatible with a smart home remote controller, researchers expect to develop SWs which function without causing any problems, even while users steer their SW within the home. Atchaya and other researchers have proposed a new framework to control different home machines using voice recognition, the Android app, and Bluetooth. This new framework is still progressing toward the perfect controller \cite{b10}.

\subsection{Other Applications}
While smart wheelchairs have traditionally been categorized as a 'niche' market, appealing primarily to a limited user base facing significant disabilities, we contend that the potential benefits of SW technologies could extend to a broader range of wheelchair users. The impact of various diseases and factors on individuals' immobility has evolved significantly over time. While the initial use of powered wheelchairs was to primarily serve individuals who were disabled due to war, their application has expanded to address a wide range of medical conditions. Conditions such as Alzheimer's disease, amyotrophic lateral sclerosis, cerebral palsy, cerebrovascular accident, multiple sclerosis, multiple system atrophy, Parkinson's Disease, progressive supranuclear palsy, and severe traumatic brain injury can result in varying degrees of immobility. Despite the relatively low prevalence of these diseases, affected individuals often face challenges in their daily mobility, necessitating assistance for even basic movements. The introduction of SWs has emerged as a promising solution, offering enhanced independence for individuals facing these conditions and reducing their reliance on external assistance. Individuals facing challenges in wheeled mobility due to impairments like ataxia, dystonia, and cognitive limitations often encounter difficulties in controlling SWs. In such cases, the development of a fully autonomous SW that is both safe and functional would greatly enhance the user mobility. This need extends to individuals with visual impairments who rely on wheeled mobility, emphasizing the broader applicability and potential impact of autonomous technology in addressing diverse mobility challenges\cite{other_applications1}.

\begin{figure}
    \centering
    \includegraphics[width=1\linewidth]{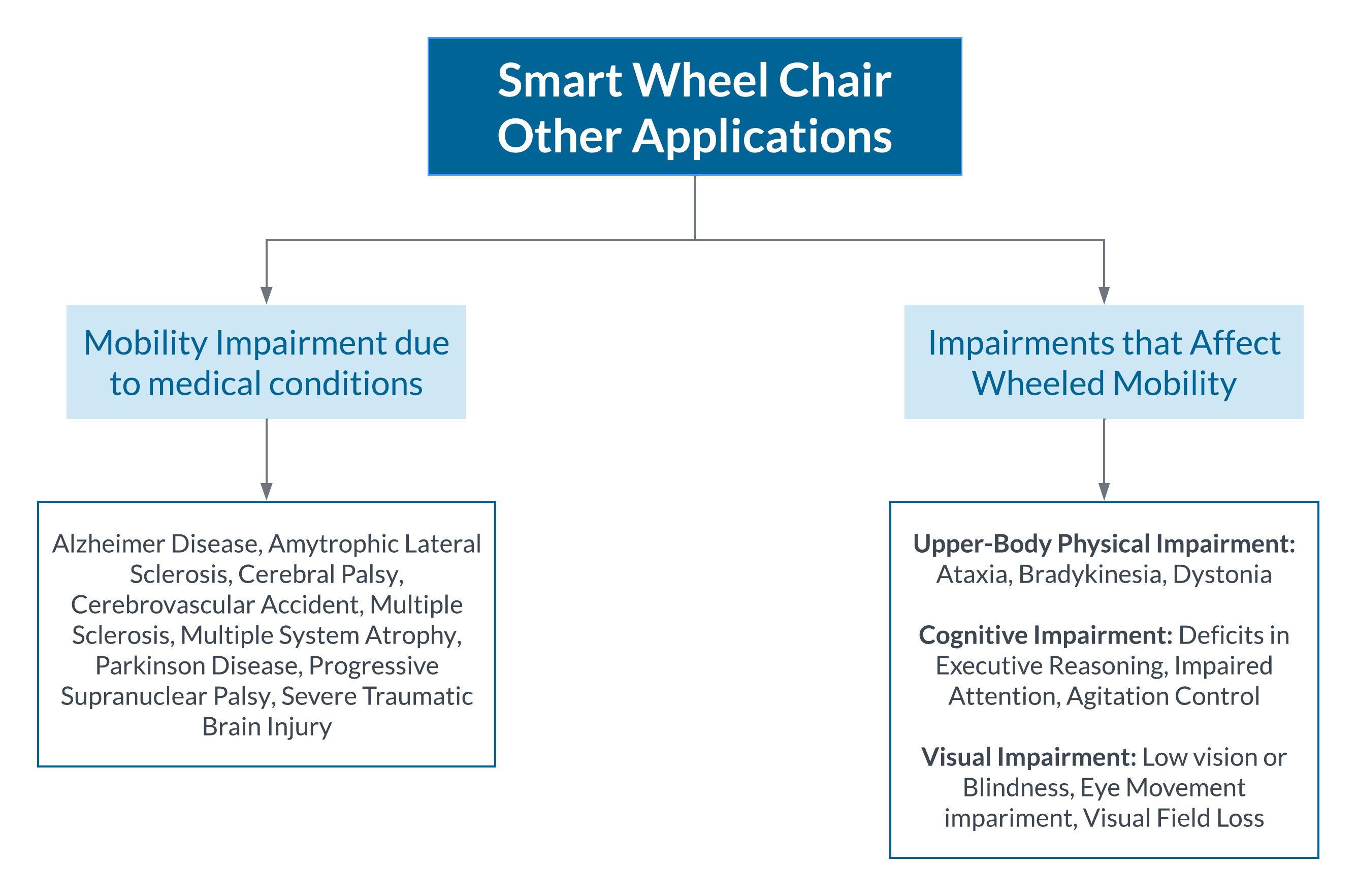}
    \caption{Smart Wheelchair - other applications with respect to medical disabilities\cite{other_applications1}}
    \label{fig:enter-label}
\end{figure}

Incorporating technological innovations into the features of SWs allows for a more diverse range of applications. One notable application is the integration of an in-house healthcare monitoring system with the SW, providing users with valuable information regarding their health and vital signs. While achieving comprehensive health monitoring may currently pose challenges, offering targeted health monitoring specific to users' respective levels of immobility presents an opportunity for further advancement in this field\cite{other_applications2}. Environments characterized by large crowds, such as airports and hospitals, pose significant challenges for SW users due to numerous obstacles. In these high-traffic scenarios, integrating features into SWs that facilitate crowd navigation and interaction with robots could prove invaluable. This enhancement aims to address the specific mobility needs of individuals with disabilities in commonly frequented public spaces.

\section{Conclusion}
This review paper provides a comprehensive analysis of smart wheelchairs (SW), highlighting their evolution, current state, and future prospects. It emphasizes the role of advanced technologies such as Brain-Computer Interface (BCI), deep learning, and various control systems in enhancing the independence and mobility of users with physical limitations. The integration of AI, sensors, and user-friendly interfaces in SWs has significantly improved their ability to navigate complex environments. Challenges such as EEG data fluctuations and the complexity of control inputs are acknowledged, along with efforts to overcome them through innovative solutions like adaptive interfaces and advanced signal processing techniques. The paper also discusses the importance of considering the emotional state and physiological needs of users, as well as the ongoing development of alternative control methods for greater convenience and efficacy. Moreover, the paper anticipates further advancements in autonomous navigation, interaction models, and the integration of SWs with smart home technologies, thereby offering promising avenues for enhancing user autonomy and quality of life. The review emphasizes the importance of fluid interaction between the user and the wheelchair, and the potential of emerging technologies to create more responsive and adaptable SW systems.

\end{document}